\documentclass{article}
\usepackage{arxivnopreprint}

\usepackage{amsmath,amsthm,amssymb}
\usepackage{adjustbox}
\usepackage{algorithm}
\usepackage{algpseudocode}
\usepackage{array}
\usepackage{csvsimple}
\usepackage{graphicx,import,color}
\usepackage{listings}
\usepackage{subcaption}
\usepackage{rotating}
\usepackage{tikz}
\usetikzlibrary{shapes,arrows,calc}
\usepackage{verbatim}
\RequirePackage{doi}
\usepackage{hyperref}
\newcommand{\reals}{\mathbb{R}}

\newcount\Comments  
\newcommand{\kibitz}[2]{\ifnum\Comments=0\textcolor{#1}{#2}\fi}

\usepackage{xcolor}

\definecolor{codegreen}{rgb}{0,0.6,0}
\definecolor{codegray}{rgb}{0.5,0.5,0.5}
\definecolor{codepurple}{rgb}{0.58,0,0.82}
\definecolor{backcolour}{rgb}{0.95,0.95,0.92}

\lstdefinestyle{mystyle}{
    backgroundcolor=\color{backcolour},   
    commentstyle=\color{codegreen},
    keywordstyle=\color{magenta},
    numberstyle=\tiny\color{codegray},
    stringstyle=\color{codepurple},
    basicstyle=\ttfamily\footnotesize,
    breakatwhitespace=false,         
    breaklines=true,                 
    captionpos=b,                    
    keepspaces=true,                 
    numbers=left,                    
    numbersep=5pt,                  
    showspaces=false,                
    showstringspaces=false,
    showtabs=false,                  
    tabsize=2
}
\begin{document}

\title{SEIR-Campus: Modeling Infectious Diseases on University Campuses}

\author{
    Matthew Zalesak \\
    School of Operations Research and Information Engineering \\
    Cornell University \\
    Ithaca, NY 14853 \\
    \texttt{mdz32@cornell.edu}\\
    \And
    Samitha Samaranayake \\
    School of Civil and Environmental Engineering \\
    Cornell University \\
    Ithaca, NY 14853 \\
    \texttt{samitha@cornell.edu}
}

\maketitle

\begin{abstract}
We introduce a Python package for modeling and studying the spread of infectious diseases using an agent-based SEIR style epidemiological model with a focus on university campuses.  This document explains the epidemiological model used in the package and gives examples highlighting the ways that the package can be used.

\center{
The package and examples can be found in the Git repository:

\url{https://github.com/MAS-Research/SEIR-Campus}
}
\end{abstract}

\section{Introduction}

SEIR-Campus is a Python package designed to aid the modeling and study of infectious diseases spreading in communities with a focus on fast computations for large university campuses. It uses an agent based framework with interactions derived from individual movement patterns and interactions. For example, in the university setting using course registration data and models of student social dynamics to simulate day-by-day spread of infections in discrete time.  Its features include:
\begin{itemize}
    \item An epidemiological model based on the SEIR model
    \item Distinction between symptomatic and asymptomatic cases
    \item Modeling interactions (e.g., course contacts and social interactions) that change day-by-day
    \item Specifying stochastic model parameters (e.g., the infectious period, recovery time etc. can be a stochastic process) 
    \item Highly customized testing protocols and quarantine procedures for individuals
    \item Integration of contact tracing
    \item Ability to add custom social networks
    \item Fast computation of large problem instances (e.g., 20,000 individuals for a semester in less than five seconds)
\end{itemize}

The package is designed to be easy to use out of the box, at a minimum only requiring formatting an input file identifying a list of individuals and a list of gatherings they attend.  The package is also designed to be highly customizable for a variety of circumstances of interest to users.  For example, the in-built infection testing model can be extended to allow for any custom function to decide who to test and what failure rates may occur.  Also, a simple default example of contact tracing is included, but any custom contact tracing algorithm can be inserted.  How customizations work, and more, is explained in the examples Section~\ref{sec:examples} to allow for simple extensions. We encourage users with implementation experience to explore the source code themselves.

Our work is motivated by the emerging literature on studying disease spread in university settings.  One such example, \cite{gressman2020simulating}, creates a model that estimates the dynamics and interactions at a university by randomly generating interactions among agents for each day, including randomization for constructing student course schedules.  Another related study~\cite{bahl2020modeling}, considers interactions in terms of agents meeting at physical locations. This is done via a graph modeling spaces on campus with agents' itineraries randomly generated as movements in the graph on an hour-by-hour basis.  As an alternative to largely randomized models, \cite{weeden2020small} uses real university transcript data to study the interaction graph of students based on historical course enrollment.  Building from these ideas, our simulator uses an agent based model that allows for dynamic interaction graph construction by integrating interaction data (e.g., from transcript data) with information/models of other social gatherings, both in a deterministic and randomized manner.  While the accuracy of the specific predictions about actual disease propagation are dependent upon the data and parameters selected by the user and other factors, the main intended purpose of this tool is to help understand the relative changes in the dynamics of the system under different assumptions.

Related to our package, Epidemics on Networks (EoN) \cite{miller2020eon} is a Python package that simulates infections in Susceptible-Infected-Susceptible (SIS) and Susceptible-Infected-Recovered (SIR) disease models both over networkx graphs and differential equation models, also including a variety of visualizations.  In contrast to EoN, our package is specifically targeted at the context of communities (e.g. university campuses) with discrete day-by-day dynamics and includes features designed to be convenient when information such as transcript data, demographic information for students, and campus group identifications are available.

One limitation of our model compared to EoN is that we only consider single, simple contagion to which agents recover with immunity.  EoN has the ability to process other simple contagion, such as infections that do not produce lasting immunity, competing diseases which cause partial cross immunity, cooperative diseases where one helps the other spread, and complex contagion where infection rates may be non-linear in the number of infectious individuals agents are exposed to.  EoN also models diseases where susceptible individuals become vaccinated at a given rate, something that could be added to our model with small modifications.  While EoN models run very fast, it is hard to compare the running time efficiency of EoN with SEIR-Campus directly due to differences in our model, as we allow daily changing interaction graphs, which means infectious periods for agents must be computed across multiple iterations to account for temporal changes in contacts.

Our package gains its computational speed by compressing all interactions within each day as an unordered collection of events.  Since the infections we are studying do not cause individuals to become contagious on the same day they are exposed, we need only to compute the groups of individuals who would interact each day and the amount of time they are exposed to each other, possibly different every day of the semester.  Events, such as individuals being quarantined or recovering from infection and no longer being susceptible or contagious, can be handled efficiently during simulation runtime.  While some stochastic elements such as infection duration, whether exposures occur, and whether exposed individuals develop symptoms are efficiently handled during runtime, other randomized elements such as changes to the daily interaction networks are preprocessed and have computation time highly dependent on the customized procedures used to create the changes.

The applications of this tool are not limited to simulations of an entire university campus; it can be used in any context in which a community of individuals whose intra-community interactions need to be modeled explicitly.  For example, professional sports leagues are keeping athletes in special ``bubbles" that are intended to isolate them from the outside world so that there are no infections internally, a context well suited for this tool.  Similarly, as many universities are moving to online-only instruction, some are considering offering special programs to keep athletes on campus.  In practice it is prudent to consider that external infections may reach such groups and simulations tools that help plan policies that are robust against the spread of infection may prove invaluable in getting the best results.

This document introduces the modeling framework we use and illustrates the software's usage through sets of examples.  The modular implementation of the simulator allows for custom choices of quarantine policy, contact tracing policies, and testing procedures, all of which are explained and included in the examples file in the package. We will use the example of student interactions in a university campus throughout the rest of the discussion, but note again that the applicability of the package is not limited to this use case.  

\section{Epidemiological Model}

This packages assumes a standard agent based Susceptible-Exposed-Infectious-Removed (SEIR) model that considers each student as an agent and models student interactions through meetings (e.g., course contacts and social interactions).  Within each meeting, all individuals have equal exposure to all other individuals for the full duration of the meeting (though we discuss generalizations in \ref{seirgeneralizations}).  SEIR models give each agent one of four states: susceptible (has not been infected with disease), exposed (has the disease but is not yet infectious), infectious (can spread disease to others, may or may not have symptoms), and removed (either recovered or otherwise unable to spread/receive disease).  An initial state is given to each agent and the purpose of the model is simulate how infections cause agents' states to change over the duration of simulation period.  While traditional agent-based SEIR models place agents as nodes in a static social network graph, either as a time-free model or with Markovian transmissions over time, this software implements the ability to have a transmission graph that changes each day, and whose weights are determined by the durations and intensities of the respective interactions, the product of which we refer to as the effective exposure time.

The key building block of our tool is meeting events, which act as an efficient shortcut for interaction graph generation by defining gatherings that recur on a regular or semiregular basis. 
Meetings can be created in many ways, e.g., by social interactions or by class interactions.  Meetings have three fundamental properties: a set of individuals, meeting dates, and meeting duration.  Classes, a key focus of our tool, can be defined in this way automatically through transcript data.  Transcripts can tell us which classes students are enrolled in, which days the classes meet, and how long they last each day. Models, perhaps aided by available data, may also be able to to create meetings that capture expected social interactions by students outside of class. Of course, students will typically have many other interactions both with each other and the community at large that are hard to capture by direct modeling.  We model these interactions via an exogenous infections rate parameter that can be calibrated appropriately.  While admittedly this is a somewhat crude approximation, our focus is on understanding the implications of the intra-campus interactions which we have the data to model (e.g., course network and specific social gatherings), and the impact of intervention policies that alter these interactions (e.g., course size restrictions, hybrid courses, limits on social gatherings). 

The SEIR model propagates infection between infectious individuals and susceptible individuals when they share a meeting, with effective minutes of exposure time as a weight for whether the disease successfully spreads each day.  This means that two individuals who share a longer class will have a higher probability of spreading an infection to each other than two individuals who share a shorter class.  While the realities of infection dynamics depend on many factors, infection rate is approximated in our model through the flexible choice of exposure duration.  For example, if students mingle before or after class, the effective duration of the class might be considered to be longer than its schedule time.  If a class has long duration but involves significant social distancing, such as some lab classes, their effective duration can be reduced to represent the duration of close contact expected during the class. In other words, the effective duration (a scaled version of the actual duration) can be used as a tool to calibrate the infectioun risk of individual meetings.  

We use a SEIR model with the following standard extensions.  First, we differentiate symptomatic infectious individuals from asymptomatic infectious individuals by creating two separate infectious states.  This distinction allows us to capture differences in infection spread rate and agent actions, such as not attending meetings after noticing the development of symptoms.  It also allows us to model compliance by treating a fraction of those who are symptomatic as asymptomatic agents, capturing behavior where  agents choose to have close contacts despite their symptoms. Second, we consider the possibility that individuals will be placed in quarantine.  The two principle distinctions with quarantine is whether the individual is actually infected, as uninfected individuals leave quarantine as susceptible individuals while infected individuals only leave upon recovery.  For the purposes of statistics, and the possibility of features that take advantage of the extra information, we create a quarantine state to mirror each of the states other than the removed state.

\subsection{Generalizations and possible future extensions}
\label{seirgeneralizations}

In the SEIR model we adopt, individuals fully interact with each other for the full duration of the meeting.  However, there are generalizations of this that might be considered for more realistic scenarios that could easily extend our model and implementation without much work.

One such extension would consider that not all individuals in meeting spend the same amount of time in class.  For example, if a small number of individuals leave early this reduction in exposure could be captured by means of a rejection process: if the person is identified as exposed in the simulation, a coin could be flipped that with some probability rejects the infection.  If multiple people have differing arrival and departure schedules this method would involve more effort to implement as it would not be compatible with the assumption that meetings cannot be subdivided, which was used to optimize transmission computation performance in the SEIR-Campus package.  In this case, within each meeting each infectious individual's transmissions are computed separately and each susceptible individual's rejection probability can be selected based on the amount of overlap between each of their schedules.

Another possible extension would consider that individuals in different part of the room have unequal chances of transmitting based on features such as physical distance.  Similar to the previous extension, this would not be compatible with the currently implemented computation optimizations since it requires infectious individuals to be considered separately from each other.  Similar to the previous case, the solution to implementing this would be to run separate computations for each infectious individual and calculate transmissions to others by coin flips weighted by a function of their physical separation based on the seating arrangement.

\section{Model Details}

Here we give a precise, detailed overview of the simulation model.  We begin by formally defining the states individuals can be in, describe the transition mechanics that govern when individuals change state, and then describe the algorithm for efficiently implementing the defined model.

\subsection{States}

Each agent in the system is in one of nine states.  The states are extensions of the four typical SEIR states, with two separate infectious states and a quarantine state to mirror each state other than removed.
\begin{enumerate}
    \item  Susceptible ($S$)
    \item  Exposed ($E$)
    \item  Infectious, asymptomatic ($I_a$)
    \item  Infectious, symptomatic ($I_s$)
    \item  Quarantined while Susceptible ($Q$)
    \item  Quarantined while Exposed ($Q_e$)
    \item  Quarantined while infectious, asymptomatic ($Q_a$)
    \item  Quarantined while infectious, symptomatic ($Q_s$)
    \item  Removed/Recovered ($R$)
\end{enumerate}

\tikzstyle{block} = [rectangle, draw, fill=blue!20, 
    text width=5em, text centered, rounded corners, minimum height=4em]
\tikzstyle{line} = [draw, -latex']
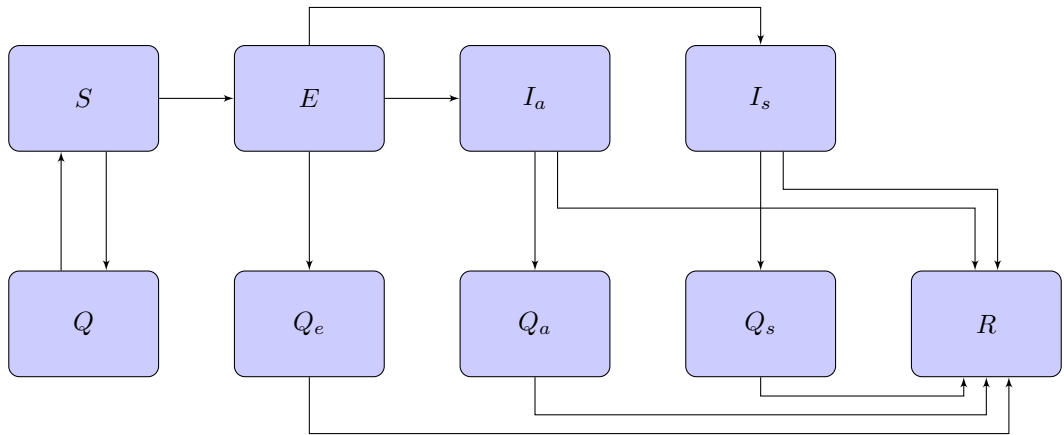
\begin{figure}
    \centering
    \begin{tikzpicture}[node distance = 3cm, auto]
        \node [block] (S) {$S$};
        \node [block, right of=S] (E) {$E$};
        \node [block, right of=E] (Ia) {$I_a$};
        \node [block, right of=Ia] (Is) {$I_s$};
        \node [block, below of=S] (Q) {$Q$};
        \node [block, below of=E] (Qe) {$Q_e$};
        \node [block, below of=Ia] (Qa) {$Q_a$};
        \node [block, below of=Is] (Qs) {$Q_s$};
        \node [block, right of=Qs] (R) {$R$};
        \path [line] (S) -- (E);
        \path [line] (E) -- (Ia);
        \path [line] (E) |- ($(E.north east) + (4,0.5)$) -| (Is);
        \path [line] ($(S.south) + (0.3, 0)$) -- ($(Q.north) + (0.3, 0)$);
        \path [line] ($(Q.north) + (-0.3, 0)$) -- ($(S.south) + (-0.3, 0)$);
        \path [line] (E) -- (Qe);
        \path [line] (Ia) -- (Qa);
        \path [line] (Is) -- (Qs);
        \path [line] ($(Ia.south) + (0.3, 0)$) |- ($(Ia.south east) + (0, -.75)$) -| ($(R.north) + (-0.15, 0)$);
        \path [line] ($(Is.south) + (0.3, 0)$) |- ($(Is.south east) + (0, -.5)$) -| ($(R.north) + (0.15, 0)$);
        \path [line] (Qe.south) |- ($(R.south west) + (0, -0.75)$) -| ($(R.south) + (0.3, 0)$);
        \path [line] (Qa.south) |- ($(R.south west) + (0, -0.5)$) -| (R);
        \path [line] (Qs.south) |- ($(R.south west) + (0, -0.25)$) -| ($(R.south) + (-0.3, 0)$);
    \end{tikzpicture}
    \caption{State flow chart.  Susceptible individuals ($S$) may be quarantined ($Q$), but then return to being susceptible afterwards.  When infected, susceptible individuals become exposed ($E$).  An exposed individual eventually becomes infectious, either asymptomatic ($I_a$) or symptomatic ($I_s$).  Infectious individuals, if not subject to any interventions, recover and enter the removed state ($R$).  Exposed ($E$) or infectious ($I_a$, $I_s$) individuals may be quarantined in ($Q_e$), ($Q_a$), or ($Q_s$) respectively.  Since these individuals are infected before they enter quarantine, they only leave once they have recovered and are moved to the removed state ($R$).}
    \label{fig:states}
\end{figure}

\subsection{Actions}

Agents in the model change state through the follow set of transition rules.  Some of the rules may not apply to all simulations, such as those that do not use testing, contact tracing, or quarantine, while other happen automatically if not interrupted, such as exposed agents becoming infectious.  Figure \ref{fig:states} gives a visual representation of agent states and possible transitions.

\begin{enumerate}
    \item \textbf{Transmission.}  Each day, infectious individuals from $I_a$ and $I_s$ can transmit infection to individuals in $S$ at a rate determined by effective exposure minutes between pairs of individuals.  This causes the state transition $S \to E$.
    \item  \textbf{Spontaneous exposure.}  Each day, individuals from $S$ become exposed spontaneously (due to infections exogenous to the interaction model) with a certain probability.  This causes them to transition $S \to E$.
    \item  \textbf{Incubation.}  Individuals in $E$ transition to either $I_a$ or $I_s$ with a given probability and transition at a rate dependent on which state they transition to. 
    \item  \textbf{Testing.}  Tests happen both at regular intervals and at scheduled times due to an individual being selected by contact tracers.  Testing latency causes results to be returned after a delay from when the test is performed, by default the next day.  Tests have certain probabilities of type 1 and type 2 errors.  A test is correctly positive when an individual is any of states $I_a$, $I_s$, $Q_a$, or $Q_s$.  When a test yields a positive result, contact tracing is invoked on the the individual who tested positive.  A positive test result causes state transitions $S \to Q$, $E \to R$, $I_a \to R$, and $I_s \to R$.
    \item  \textbf{Contact tracing.}  Contact tracing on an individual identifies a group of peers that have been near the individual for an extended period of time, though the precise method used to do this is not part of the model specification.  Each individual identified in contact tracing is moved into quarantine for a specified number of days.  It causes the state transitions $S \to Q$, $E \to Q_e$, $I_a$ to $Q_a$, and $I_s \to Q_s$. Each individual quarantined may be tested
    \item  \textbf{Release from Quarantine.}  It is assumed that exposed individuals in quarantine are detected at some point during the infection and never leave until they are recovered.  Thus, they leave quarantine at the later of their recovery time and the expiration of their Quarantine.  This results in state transitions $Q_e \to R$, $Q_a \to R$, and $Q_s \to R$.  Additionally, unexposed individuals return to the susceptible state with the transition $Q \to S$.
    \item  \textbf{Recovery.}  On an individual's recovery date, they are no longer infectious and move to the removed state, through the transition $I_a \to R$ and $I_s \to R$.
\end{enumerate}

\subsection{Internal Structure}

The model is executed over a set of days $D$, which includes weekends and holidays from the beginning of the semester.  The model contains a bipartite graph between individuals and meetings $M$.  Since there are courses at the university that have irregular meeting days and times, there is a function that gives the duration the course meets each day $Dur : (D, M) \to \reals_+$.  There are sets corresponding to each of the 9 states each individual can be in.  In addition, each member of $E$, $I_a$, $I_s$, $Q$, $Q_a$, and $Q_s$ are associated with a date at which they will automatically transition to a subsequent state if there are no interventions.  These transition times are stored in associative array data structures.

The only two types of actions that are not scheduled are contact tracing and testing.  While contact tracing is typically scheduled on the same day that a positive test result is returned and therefore is not scheduled in advance, testing can happen under a variety of conditions that may require prior scheduling.  For example, weekly infection testing may occur or someone who is quarantined for any reason may be scheduled to receive a test result the following day.  In general, an individual may be tested multiple times if they are tested while susceptible or exposed but not yet contagious.  However, only in the case of a susceptible individual does leaving quarantine mean returning to the general population.  Therefore, in effect only susceptible individuals can ever test positive more than once and only susceptible individuals can invoke contact tracing multiple times.

\subsection{Order of Operations}

The simulation processes time in units of days\footnote{Using units of days is sufficient and without loss when infections cause agents to change state on the order of multiple days.}.  For each simulated day, events are processed in the following order:
\begin{enumerate}
    \item  Scheduled transitions due that day (such as an individual recovering) are processed and sets are updated.
    \item  Testing results come in.  A list is made for contact tracing and quarantines are updated.
    \item  Contact tracing is performed and quarantines are updated.
    \item  Daily infection spread is processed.
\end{enumerate}

\subsection{Processing Daily Infections}

In traditional SEIR models on static interaction graphs, all of the outgoing edges from an infectious individual to neighboring susceptible individuals are considered and a weighted coin is flipped to determine whether the infection spreads to each susceptible individual.  The coin flips can even be preprocessed on an edge-by-edge basis.  However, this approach is not possible for models where the interaction graphs change dynamically.  Rather, we use a procedure that aims to efficiently compute transmission by considering each day sequentially.

In simulating the model each day, infection computations are done on a meeting-by-meeting basis.  Each of the meetings is processed sequentially to determine that day's transmissions.  First, the set of infectious individuals are identified in each meeting.  Then, since infectious individuals transmit infections to susceptible individuals at a fixed rate per effective minute of exposure time, and since we assume that meetings are fully connected interactions, we next determine the total number of exposure minutes experienced by susceptible individuals in the meeting that day.  Exposure per susceptible individual is equal to the total number of infectious individuals multiplied by the transmission rate and effective duration of the meeting.  To determine which susceptible individuals become infected, we use one of two procedures depending on which involves fewer calculations.

If the effective exposure per susceptible individual is less than one, then the expected number of transmissions during that class is less than the number of susceptible individuals.  In this case we use a Poisson process to determine the total number of infections that occur and then uniformly, with replacement, draw that number of individuals from the susceptible population.  On the other hand, if the number of expected transmissions is greater than the number of susceptible individuals we instead use a binomial random variable for each susceptible individual using the probability per individual of transmission.

\subsection{Time Considerations}

The running time of each repetition of the model is divided into two portions: preprocessing, which creates the set of nominal interactions, and runtime, which processes infection spread and agent states.  While the runtime of the preprocessing step cannot be stated since it allows an arbitrary number of customizations, the largest driving factors of runtime for the simulation are the number of individuals who eventually become infected and the number of distinct meetings in the simulation.  The influence of these factors comes directly from the model: each day, the spread of infection is event driven and carries out a computation based on the set of meetings that have at least one infected individual.

It is worth mentioning some alternative approaches to processing infection spread that we rejected either due to runtime considerations or compatibility problems with our dynamic model.  Since classical SEIR models compute the spread of infection by flipping coins on weighted edges between infectious and susceptible individuals, a natural first idea is to compute a large static graph of contacts for each infectious individual that covers all susceptible individuals they will contact during their infectious period, complete with the appropriate weights based on the the sum of the contact times from each days' meetings.  This approach fails, however, to allow generalizations with testing and contact tracing procedures which may ultimately limit the number of days the infectious or susceptible individuals have contact with others.  We decided instead to compute the spread of infection on a day-by-day basis.  Computationally, this is just a reordering of events since in either case we must determine the total contact minutes between pairs of individuals on each day's interaction graph.  Further, this can sometimes lead to simplifications since some individuals may become infected or quarantined on later days, reducing the number of computations needed.

Another key note is the choice to have two procedures to compute transmission based on the expected number of individuals to infect.  Initially we anticipated that a very small number of people would become infected in each meeting each day, suggesting that the most efficient manner to compute infections would be to compute the number of susceptible individuals that would become infected using a Poisson process and then drawing, with replacement, that number of people from the susceptible population and label them as infected.  In scenarios where few people become infected this procedure can be implemented very fast in Python since it can be performed by external libraries, such as Numpy's \verb|random.choice| function, which offer significant speed advantages over explicitly written \verb|for|-loops.  However, in scenarios where many individuals are infectious it becomes slow.  In these cases, the expected number of infections in a meeting may exceed the number of susceptible individuals, which can lead the \verb|choice| function to redundant work since we draw with replacement.  Therefore, in these situations, we instead use an alternate procedure of flipping coins for each susceptible individual using an explicit \verb|for|-loop, which ensures no redundant work is done.

\section{Usages Examples}
\label{sec:examples}

The best way illustrate the flexibility of the model is through a series of demonstrations.  This section explains the basics for loading data into the model, and then discusses ways the model can be expanded and customized.  Since the SEIR-Campus model is implemented in Python, each of the examples below is also written in Python and is included in the \verb|Examples| Jupyter notebook from the git repository.  To get the most value, the reader is encouraged to play along with these examples as they read.

\subsection{Basic Usage}

To begin, we will create and load a simple example to demonstrate how the simulation works.  We will describe the input file format, how the data is loaded, and how repetitions of the simulation are performed.  This model will assume that individuals only interact through course contacts and that no testing or contact tracing programs are in place.

\subsubsection{Input file format.}

A course description file has three types of entries: student entries (demographic information), meeting entries (used for classes or precomputed social gatherings), and group entries (used for constructing custom social groups).  Each line of the file contains a single entry and is in the form of a Python dictionary, whose key is the entry type and whose value contains a dictionary of the appropriate information for each entry type.  For example, a stem student enrolled as an undergraduate might have an entry:
\begin{lstlisting}
{'student': {'id': 12345, 'demographics': {'gender': 'm', 'is_grad': 'n', 'field': 'STEM'}}}
\end{lstlisting}
The key \verb|student| indicates that this entry is for a student, rather than a meeting or group.  The value is a dictionary which defines the student \verb|id| and a demographics dictionary.  The demographics dictionary has no fixed format; it is a convenience for use in building custom groups and meetings by user code.

Meetings, used to build classes or other fixed social gatherings, are defined with the key \verb|meeting|.  As an example, a class might be created using
\begin{lstlisting}
{'meeting': {'id': '5166', 'info': {'name': 'CLASS 5166'}, 'meets': [('9/2/2020', '11/13/2020', 'MWF', 60)], 'members': [1452, 1633, 1960, 2305, 2417, 2456, 2535, 2635, 2682, 2738, 2767, 2832, 2873, 2945, 2975, 3235, 3250, 3332, 3334, 3644, 3812, 3835, 4036, 4053, 4059, 4154, 4220, 4288, 4423, 4582, 4591, 4848, 4925, 4932, 4976, 5033, 5143, 5230, 5251, 5258, 5261, 5274, 5365, 5379, 5422, 5455, 5463, 5474, 5514, 5516, 5523, 5600, 5634, 5644, 5663, 5777, 5802, 5887, 5915, 5944, 5957, 6026, 6061, 6071, 6095, 6145, 6401, 6446, 6624, 6706, 6734, 6786, 6873, 6898, 6978, 7092, 7201, 7408, 7448, 7462, 7595, 7601, 7666, 7751, 7771, 7772, 7809, 7818, 7832, 7890, 7987, 7998, 8049, 8102, 8114, 8125, 8300, 8316, 8339, 8371, 8376, 8386, 8399, 8446, 8449, 8473, 8496, 8520, 8545, 8613, 8616, 8620, 8630, 8631, 8645, 8725, 8726, 8739, 8804, 8826, 8854, 8875, 8943, 8956, 8966, 8971, 8972, 9000, 9034, 9036, 9062, 9141, 9212, 9221, 9288, 9310, 9311, 9342, 9492, 9500, 9535, 9727, 9860, 9880, 9975, 10108, 10139, 10147, 10208, 10209, 10304, 10311, 10321, 10362, 10413, 10437, 10545, 10618, 10667, 10716, 10791, 10806, 10853, 10895, 11054, 11236, 11260, 11402, 11435, 11553, 11594, 11604, 11653, 11663, 12346, 12511, 12933, 13193, 13287, 13449, 13572, 13713, 13970, 14041, 14243, 14287, 15000, 15110, 15419, 15522, 15657]}}
\end{lstlisting}
The meeting is defined by an \verb|id|, a list of \verb|members| comprised of student ids, and an \verb|info| dictionary.  The \verb|info| dictionary contains a list of \verb|name|s the class is referred to (typically one but classes are sometimes cross listed) and \verb|meets| information which lists tuples of the form \verb|(start date, end date, weekday pattern, duration)|.  Dates are written using Month/Day/Year convention and weekday patterns can contain any combination of the letters \verb|MTWRFSU|.  Many classes need only one such tuple, but some classes with custom schedules may require multiple.

Finally, there are group definitions.  These are not required, but provide an easy way to load information about groups into the system for later manipulation into social meetings. Groups are not the same as classes or social meetings, per se.  The intention is that they are for creating meetings dynamically, such as the meeting patterns of members of a club which may be random or meetings that the user wants to create in custom ways.  This is in contrast to typical meeting definitions which are static and computed at the time of the creation of the data set.

A group only defines a name and a list of members.  An example entry might be,
\begin{lstlisting}
{'group': {'name': "Women's Varsity Ice Hockey", 'members': [311, 351, 811, 1151, 1251, 1331, 1391, 1411, 1471, 1571, 1631, 1851, 1871, 1931, 1951, 1991, 2031, 2051, 2071, 2111, 2131, 2151, 2171, 2191, 2211]}}
\end{lstlisting}

\subsubsection{Production of sample data}

The data included in the repository for this package is produced from a combination of publicly available data and a randomized model for courses and students.  The underlying course network is from the network used by Weeden and Cornwell \cite{weeden2020small} in the supplemental materials \cite{weeden_data}.  From this data, student demographic information is assigned by labeling all even student id numbers as female and all odd student id numbers as male, something that will be used to create varsity teams later on.  Graduate/undergraduate status and field of study are directly from the Weeden and Cornwell data.

Since the data has been anonymized and therefore does not contain any information about the identifies of the courses, we must make artificial choices about meetings times and dates.  Despite the wide variety in class meeting times in reality, we assume all classes in the data set meet for 60 minutes.  We determine the weekly meeting pattern randomly, as in \cite{gressman2020simulating}, by assuming 40\% of courses meet Monday/Wednesday/Friday, 20\% meet Monday/Wednesday, and 40\% meeting Tuesday/Thursday.  All starting and end dates are selected to be September 2, 2020 through November 13, 2020, in order to align with the Cornell University Fall 2020 academic calendar.  We also note that October 14, 2020 is the only holiday during the semester, though holiday information is added during runtime and is not part of the data set.

For varsity data, we synthetically create the teams as follows.  First we chose four team names and roster sizes as:
\begin{itemize}
    \item  Women's varsity Ice Hockey, 25 students
    \item  Men's varsity Squash, 15 students
    \item  Varsity Football, 100 students
    \item  Varsity Gymnastics, 20 students
\end{itemize}
Then we select students, without replacement, from the list of students provided in the Weeden and Cornwell data.  We do this by starting with student index 10 for women and 11 for men, and incrementing the indices 10 at a time, selecting a student subject to not being labeled as a graduate student.  After populating a team, the starting index for populating the next team resumes where the previous index left off.

The resulting list of students, courses, and varsity teams is then formatted appropriately as an input file for the simulator.  Remember, this data set is not intended to be a realistic scenario; rather, it is just a random data set to show the functionality of the SEIR-Campus model.

\subsubsection{Loading the data.}

The basic object of the simulation is the \verb|Semester| object.  This object stores all fields described in the input file, expands the meeting schedules into daily meeting lists, and accepts modifications during runtime.  To create the \verb|Semester| object, we pass the data filename and a list of holidays that classes will not occur on to the object constructor.  The \verb|Semester| object will automatically not expand class meetings on those days.

\begin{lstlisting}
from datetime import datetime

holiday_list = [(2020,10,14),]
holidays = set(datetime(*h) for h in holiday_list)
semester_full = Semester('publicdata.data', holidays)
\end{lstlisting}

\subsubsection{Running the simulation}

Creating a simulation is now very simple.  We create a \verb|Parameters| object using the default settings, but changing the number of repetitions to 100 (note that repetitions is the only, and optional, argument to the \verb|Parameters| object; however we will discuss all of its members as we go through the examples).  Then we use the convenience function \verb|run_repetitions| to execute multiple instances of the simulator.

\begin{figure}
    \centering
    \includegraphics[width=\textwidth]{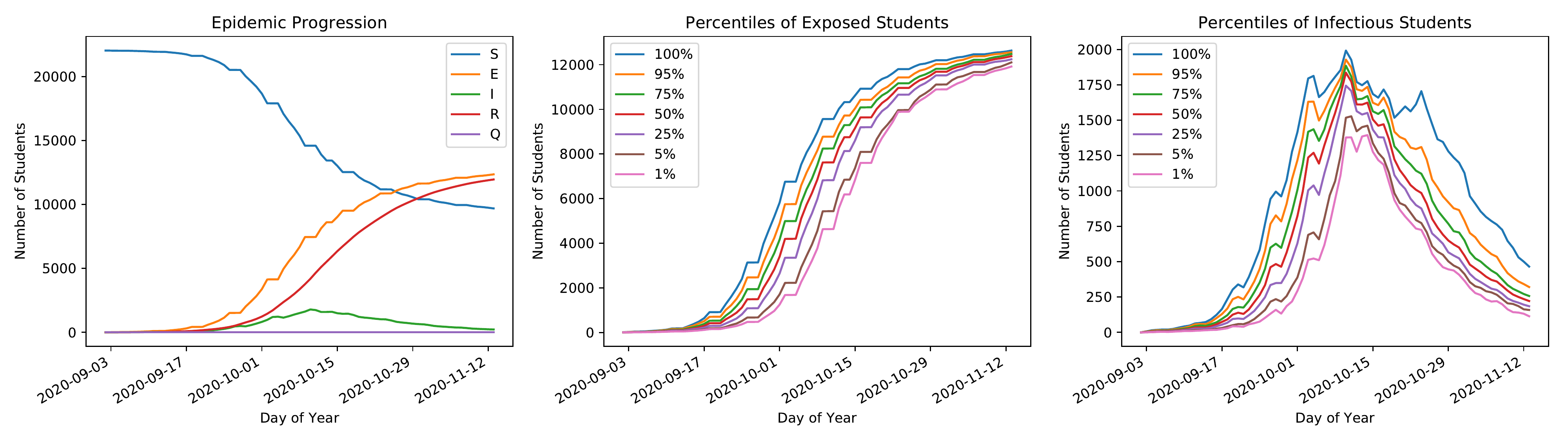}
    \caption{Results from running a basic simulation.  Each simulation outputs three charts.  The first shows the number of susceptible, exposed, infectious, recovered, and quarantined individuals by date.  The second shows the percentiles for the number of exposed individuals by date across all repetitions.  The final chart shows the percentiles for the number of infectious individuals by date across all repetitions.}
    \label{fig:basic_simulation}
\end{figure}

\begin{lstlisting}
parameters = Parameters(reps = 100) # <-- Can change from default number of repetitions
run_repetitions(semester, parameters)
\end{lstlisting}

Refer to Figure \ref{fig:basic_simulation} for sample results.

\subsection{Asymptomatic and Symptomatic Individuals}

SEIR-Campus allows you to make distinctions between individuals who are asymptomatic and those who are symptomatic.  The primary distinction we make is that symptomatic individuals are 1) more contagious than asymptomatic individuals and 2) will self report to take a test when they notice symptoms developing.  Control over which individuals are symptomatic and how long infections last can be fully customized, though some convenient defaults are also provided.  For full control, you must make a class that provides the following function call:
\begin{lstlisting}
def duration(self, simulation, student, date):
    # (Custom code goes here)
    return infectious_start_date, infectious_end_date, is_symptomatic
\end{lstlisting}
The function is called by the simulator with a reference to itself, as well as the student's id and the date of infection.  The return value is the day the individual will become infectious, the day the individual will no longer be infectious (possibly because they isolate themselves after recognizing symptoms), and a boolean indicating whether the individual will develop symptoms. For users who don't need as much flexibility, there are two built-in classes ready to use.  If you want control over the infection duration without considering symptomatic individuals, you can use the \verb|BasicInfectionDuration| class.  It takes two arguments: the rate at which individuals become contagious (per day) and the rate at which contagious individuals recover (in days).  It then draws actual durations from a geometric distribution with the given parameters.
\begin{lstlisting}
duration_computer = BasicInfectionDuration(1 / 3.5, 1 / 4.5)
\end{lstlisting}
This gives a mean duration of $3.5$ days until an infected individual becomes infectious, and an average of $4.5$ days subsequently until they recover.

If symptomatic individuals are of concern, you can use the class \verb|VariedResponse|.  This class is initialized with the rate at which individuals become contagious, the rate at which individuals recover, the rate at which individuals develop symptoms (if they will be symptomatic), and the percentage of individuals that never develop symptoms.  The default used by the simulator is:
\begin{lstlisting}
duration_computer = VariedResponse(1 / 3.5, 1 / 4.5, 1 / 2, 0.75)
\end{lstlisting}
This gives the same number of days, on average, as before until an individual becomes infectious and then recovers, but also provides that an average of 2 days pass after becoming infectious before the individual starts showing symptoms and that 75\% of the individuals are asymptomatic.  This is also the default method used by the simulator to compute infection durations if you do not provide an alternative.

The infection duration method of your choice can be passed to the simulator by placing it in the \verb|Parameters| object.  Here is an example where all individuals are asymptomatic.  Note that the code for functions that are already implemented can be seen in the code repository and can serve as a template for creating new ones.

\begin{figure}
    \centering
    \includegraphics[width=\textwidth]{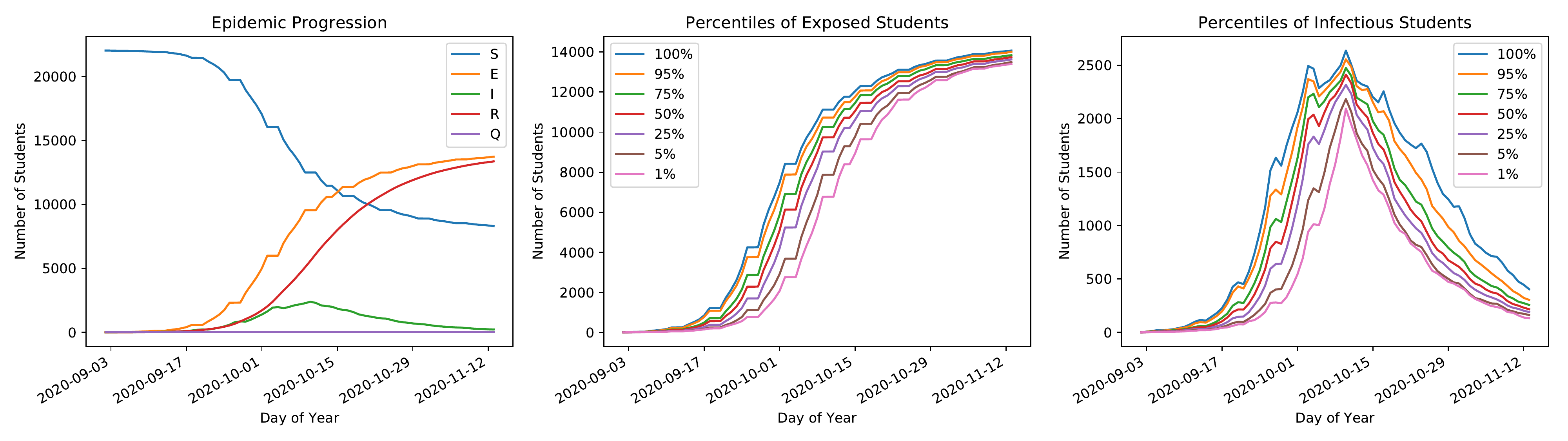}
    \caption{The three output charts for the basic simulation with the modification that none of the individuals report/have symptoms.}
    \label{fig:asymptomatic}
\end{figure}

\begin{lstlisting}
parameters = Parameters(reps = 100)
parameters.infection_duration = BasicInfectionDuration(1 / 3.5, 1 / 4.5)
run_repetitions(semester, parameters)
\end{lstlisting}
Sample results for this code are shown in Figure \ref{fig:asymptomatic}.

\subsection{Testing}

The first intervention strategy we will look at is testing, which identifies whether individuals are infected.  In simulation, determining whether an individual is infected comes from inspection of the individual's state.  $E$, $I_a$, and $I_s$ could all be states that might cause a positive test.  Some infections may only be detectable while the individual in infectious, in states $I_a$ and $I_s$.  The goal of a testing policy is to report individuals who are infected so that they can quarantine to prevent further spread of infection and possibly help with contact tracing efforts.  While the default setting in the simulation is to not use any testing, there are multiple interfaces one can use.

The \verb|Simulator| objects tracks several quantities across the duration of the simulation to help facilitate testing policies, even though the \verb|Simulator| itself leaves testing to outside functions.  The most useful quantity for knowing who to test is a dictionary called \verb|test_requests|, indexed by date.  This dictionary contains a list for each day of individuals who have been recommended for testing on that day.  Individuals may have been added to this list for several reasons, for example individuals who show symptoms that may wish to be tested or individuals reported by contact tracing for testing.  Of course, individuals not on the \verb|test_request| list might also be tested as well, especially in infection surveillance programs that purposely test individuals that might not know they need to be tested.  Other information tracked by the \verb|Simulator| is also directly relevant to determining test results, such as state information that tracks which individuals are currently infectious.

The \verb|Simulator| object offers a very general way to implement testing policies by placing the responsibilities for testing in a callable object given by the user and placed in the \verb|Parameters| object.  The goal of the object is to report to the \verb|Simulator| which individuals have taken a test that will return a positive result.  The object is called by the simulator on each simulated day.  At a more detailed level, the object should be an instance of a class that provides the function \verb|testing| whose arguments take a reference to the \verb|Simulator| and the current date as arguments.  The \verb|testing| function can determine who to test and who will get a positive result using any of the resources mentioned in the previous paragraph, possibly even including random elements if false positive and false negatives are to be modeled.  Positive results are reported through a function exposed by the \verb|Simulator| object called \verb|schedule_positive_test_result|, which takes as an argument the id of the individual who tests positive.  It will automatically notify the individual of their test result after the default test return latency (1 day).

As an example, consider the following testing policy, which is included in the SEIR-Campus package.

\begin{lstlisting}
class IpGeneralTesting:
    def __init__(self, test_groups):
        self.test_groups = test_groups
    def testing(self, simulation, date):
        to_test = set(self.test_groups[date.weekday()])
        to_test.update(simulation.test_requests[date])
        for s in to_test:
            if s in simulation.state.Ia or s in simulation.state.Is or \
                    s in simulation.state.Qa or s in simulation.state.Qs:
                simulation.schedule_positive_test_result(s)
\end{lstlisting}

This testing policy tests a fixed group of individuals each day of the week, as well as any who have been identified by contact tracing.  The class initializes with a dictionary mapping the weekday (as a number 0-6).  When the \verb|testing| function is called, it gets the set of individuals that regularly test on that day of the week and adds to that the set of individuals who need to be tested because of contact tracing or showing symptoms.  For each individual identified, if they are infectious or in quarantine and infectious it reports they have tested positive.

The flexibility offered by creating a testing class allows potential for other considerations, such as testing that does not conform to weekly patterns, false negative and false positive test results, and failure to participate for individuals.

SEIR-Campus comes with some convenient ready-made testing policies to explore.  Below are descriptions and how to create them.

\begin{itemize}
    \item \verb|IpWeeklyTesting(semester, weekday = 0)| Test all individuals on a given day of the week (default is Monday).
    
    \item \verb|IpWeekdayTesting(semester)|  Test individuals once per week, with individuals evenly assigned to be tested on Monday - Friday.
    
    \item \verb|IpDailyTesting(semester)|  Test individuals once per week, with individuals evenly assigned to be tested each of the seven days of the week.
    
    \item \verb|IpRollingTesting(semester, days = 5)|  Test all individuals on a 5 day cycle, independent of alignment with the week.  Individuals evenly assigned to test each day of the rolling cycle.
\end{itemize}

\begin{figure}
    \centering
    \includegraphics[width=\textwidth]{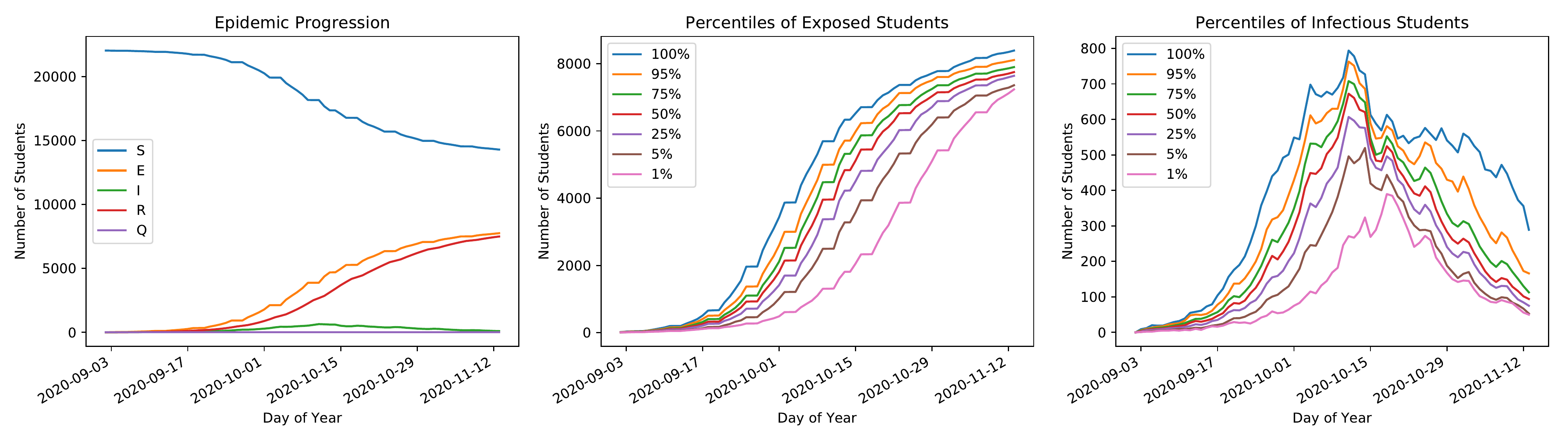}
    \caption{The three output charts for the basic simulation with the modification that each weekday, $1/5$ of the student body is tested for infection.  Those that test positive are placed in quarantine the next day when the result comes back.}
    \label{fig:weekday_testing}
\end{figure}

Here we run an example of testing, with sample results shown in Figure \ref{fig:weekday_testing}.
\begin{lstlisting}
parameters = Parameters(reps = 100)
parameters.intervention_policy = IpWeekdayTesting(semester)
run_repetitions(semester, parameters)
\end{lstlisting}

\subsection{Contact Tracing}

Contact tracing is an essential tool in public health.  Contact tracing allows those who may have been exposed to an infectious individual to either be tested or quarantined before they would normally be tested or develop symptoms.  The default setting in SEIR-Campus is to not perform any contact tracing, though the package comes with some simple policies that can be used as well as the ability to make custom policies.  A custom contact tracing policy can be made by producing a class that provides the function \verb|trace|, taking a reference to the \verb|Simulator| object and the date as arguments.  To illustrate this, let's look at the implementation of \verb|BasicContactTracing| in SEIR-Campus.

\begin{lstlisting}
class BasicContactTracing:
    def __init__(self, quarantine_length, trace_length = 3):
        self.quarantine_length = quarantine_length
        self.trace_length = trace_length
    def trace(self, simulation, date):
        ''' For each student to trace, find all of their peers from the past
            3 days.  Then, quarantine all those peers for some days and
            schedule a test for them. '''
        trace_dates = [date - timedelta(days = d + 1) for d in range(self.trace_length)]
        for s in simulation.contact_trace_request[date]:
            recent_meets = set()
            for m in simulation.semester.student_enrollment[s]:
                if simulation.semester.meeting_type[m] != MeetType.UNTRACEABLE:
                    for d in trace_dates:
                        if m in simulation.semester.meeting_dates[d]:
                            recent_meets.add(m)
                            break
            peers = set()
            for m in recent_meets:
                peers.update(simulation.semester.meeting_enrollment[m])
            if len(peers):
                peers.remove(s)
                for peer in peers:
                    simulation.initiate_quarantine(peer, date, self.quarantine_length)
\end{lstlisting}

The \verb|BasicContactTracing| class gets the contact tracing list from the \verb|Simulator| object's dictionary member \verb|contact_trace_request|, indexed by date.  By default, this policy looks back three days at all meetings for the individual that are traceable and sends all members of those groups to quarantine.  It does this using the \verb|Simulator| object's member function \verb|initiate_quarantine|, which requires arguments of the student's id, the date the quarantine starts, and optionally the length of the quarantine.  If no length is given, the default specified in the \verb|Parameters| object is used.

\begin{figure}
    \centering
    \includegraphics[width=\textwidth]{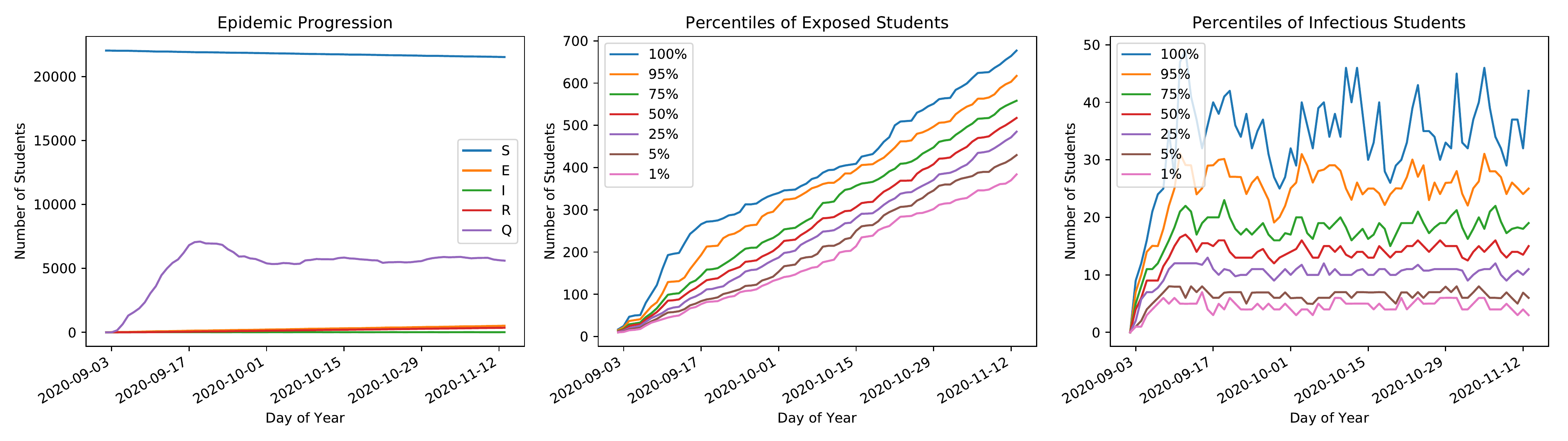}
    \caption{The three output charts for the basic simulation with the modification that when individuals show symptoms and take a test with a positive result, everyone they have been near in the past three days is placed in quarantine.}    \label{fig:contact_tracing}
\end{figure}

Here we demonstrate the usage of the basic form of contact tracing.
\begin{lstlisting}
parameters = Parameters(reps = 100)
parameters.contact_tracing = BasicContactTracing(14)
run_repetitions(semester, parameters)
\end{lstlisting}
Sample results from this code are shown in Figure \ref{fig:contact_tracing}.

\subsection{Exploring Alternative Course Designs}

In preparing a policy for operation, course design can have a big impact since it is a primary contributor of social interactions under the control of the university.  The \verb|Semester| object in SEIR-Campus comes with function to help explore modifications to the course meeting design by adding and removing individual courses.  There are also several built in functions that can modify a semester in specific ways.

To add a course, the \verb|Semester| object provides a function \verb|add_meeting|.  It takes as arguments

\begin{enumerate}
    \item  \verb|name| The id the course will be referenced by in the simulation.
    \item  \verb|info| A dictionary of information about the meeting.
    \item  \verb|meets| Can be given in one of two forms.  The first form is a dictionary, indexed by date, that gives the duration of the class meeting on the days it meets.  The second form is a list of tuples.  Each tuple contains a start date, an end date, a weekday meeting pattern, and a duration.
    \item  \verb|members| A list or set of student ids corresponding to individuals enrolled in the course.
    \item  \verb|meet_type| One of the following: \verb|MeetType.COURSE|, \verb|MeetType.SOCIAL|, or \verb|MeetType.UNTRACEABLE|.
    \item  \verb|holidays| If giving the \verb|meets| argument as a list, this is a list of dates to exclude from class meetings.
\end{enumerate}

To remove a course, you only need to know the name (id) of the course that should be removed, using the function \verb|remove_meeting|, taking the name as the only argument.  There is also a function, \verb|clean_student_list|, that removes any individuals that are not part of any meeting.

The usage of these function can be shown through the following example.  Suppose we want to move all classes with capacity higher than 50 online, meaning they no longer represent in-person contact.  The following built-in function accomplishes this:
\begin{lstlisting}
def make_alternate_smallclasses(semester_base, max_size = 50):
    ''' No large course: only classes with at most 50 students. '''
    semester = copy.deepcopy(semester_base)
    canceled_meetings = [m for m, ss in semester.meeting_enrollment.items() 
                        if len(ss) > max_size 
                        and semester.meeting_type[m] == MeetType.COURSE]
    for m in canceled_meetings:
        semester.remove_meeting(m)
    return semester

semester_alt_small = make_alternate_smallclasses(semester, max_size = 50)
\end{lstlisting}

First this class makes a clean copy of the \verb|Semester| object.  Then it makes a list of all courses with capacity greater than 50.  Each class in this set is then removed from the meeting list through the \verb|remove_meeting| function provided by the \verb|Semester| object.

SEIR-Campus also comes with two other functions to modify schedules.  The first is \verb|make_alternate_hybrid|, which randomly separates courses into two groups and then, alternating each week, schedules one group to be in-person and the other to be online.  Online education does not involve contact, so the class is effectively canceled on that week.  The second is \verb|make_alternate_splitclass|.  This function splits all courses in half (effectively into two classes each with half of the students) and then alternately schedules, week by week, the two splits.  As an example, the following code creates and simulates a hybrid schedule starting from the loaded data.  

\begin{figure}
    \centering
    \includegraphics[width=\textwidth]{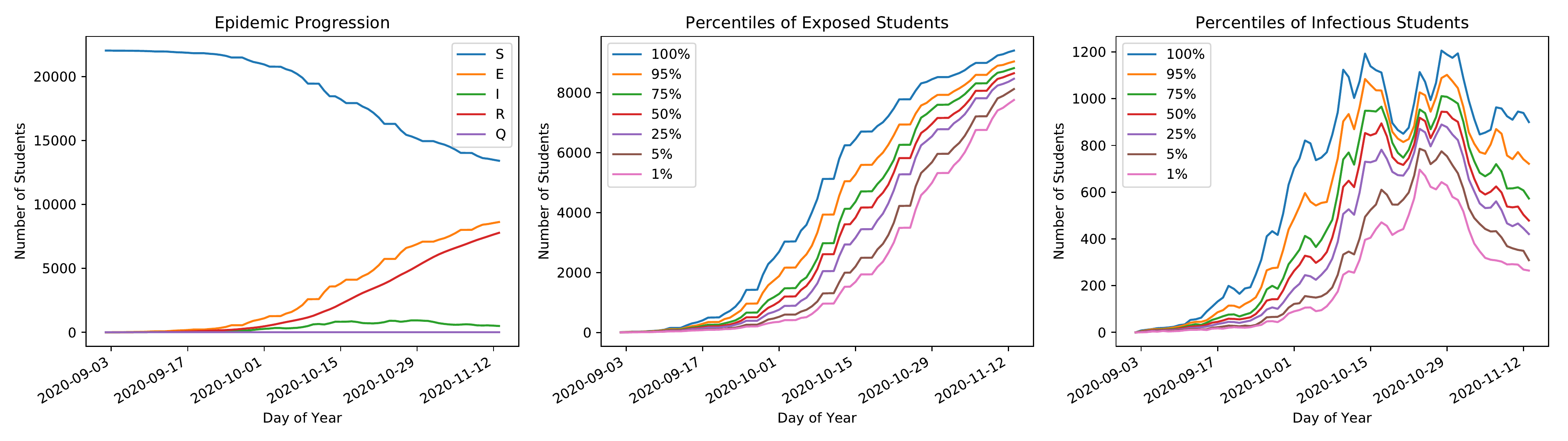}
    \caption{The three output charts for the basic simulation with the modification that courses are split into two groups, each groups alternating each week between meeting in-person and meeting online.}
    \label{fig:hybrid_class}
\end{figure}

\begin{lstlisting}
semester_alt_hybrid = make_alternate_hybrid(semester)
parameters = Parameters(reps = 100)
run_repetitions(semester_alt_hybrid, parameters)
\end{lstlisting}
Sample results are shown in Figure \ref{fig:hybrid_class}.

\subsection{Social Groups}

While social groups can have a large impact on spread of infectious diseases, unlike courses the precise interaction network is not known.  The SEIR-Campus \verb|meeting| objects provide a framework to add social interactions by modeling these interactions as courses with specific meeting durations and dates.  For example, a group of friends that meets every weekend can be modeled as a class that meets for an hour or two on Saturdays.  In practice, social networks are often not known exactly and instead generative models are used to create interaction networks.  The social network features in SEIR-Campus come with an convenient type of object, called a \verb|Cluster|, that is specifically intended for use in producing social networks.  A \verb|Cluster| is like a simplified form of course, only containing a list of members and a dictionary, indexed by date, of meeting durations.  The sample social-network generators all return lists of \verb|Cluster|s.  Once all desired social interactions have been produced, the list of \verb|Cluster|s is passed to the function \verb|make_from_clusters| which automatically expands each \verb|Cluster| into a social meeting in the \verb|Semester| object.

\subsubsection{Randomized social meetings.}

The simplest way to produce social meetings is to produce them randomly from the population.  The built-in function \verb|make_randomized_clusters| provides an easy way to do this, given the population and a \verb|ClusterSettings| object.  This object is specifically tailored for use in the built-in functions for making clusters but is not generally necessary if you wish to make you own custom cluster generators.  The built-in functions work on the following principle: from the population, draw a certain number of clusters of a certain size (to the extent possible).  Then, each individual interacts with everyone in their cluster for a predefined amount of time, given that the day is either a weekday or a weekend.  To enable this, the \verb|ClusterSettings| object contains the following settings:
\begin{itemize}
    \item  \verb|start_date|, \verb|end_date| Time range for the set clusters
    \item  \verb|weekend_group_count|, \verb|weekday_group_count| Number of individuals per group, by weekend or weekday.
    \item  \verb|weekend_group_size|, \verb|weekday_group_size| Number of individuals to draw from each group each day to form a cluster, by weekend or weekday.
    \item  \verb|weekend_group_time|, \verb|weekday_group_time| Amount of time individuals in each cluster spend with each other.
\end{itemize}

A simple way to produce a social network for a \verb|Semester| object is illustrated below.  In this example, each weekday 280 groups of 10 random students meet for 120 minutes and each day of the weekend 210 groups of 20 random students meet for 180 minutes.
\begin{lstlisting}
settings = ClusterSettings(
        start_date = min(semester.meeting_dates), end_date = max(semester.meeting_dates),
        weekday_group_count = 280, weekday_group_size = 10, weekday_group_time = 120,
        weekend_group_count = 210, weekend_group_size = 20, weekend_group_time = 180)
clusters = make_randomized_clusters(semester.students, settings)
final_semester = make_from_clusters(semester, clusters)
\end{lstlisting}
First we create a settings object, create the clusters using the \verb|make_randomized_clusters| function, and then apply them to the \verb|Semester| object using the \verb|make_from_clusters| function.

Another built in function is \verb|make_social_groups_pairs|.  This function, which as an argument takes a \verb|Semester| object, a number \verb|fraction_paired| for the percentage of individuals that are paired, a number \verb|interaction_time| for the number of minutes each pair interacts per day, and an optional set \verb|excluded| of students ids of individuals that should not be considered in the pairing process.

\begin{figure}
    \centering
    \includegraphics[width=\textwidth]{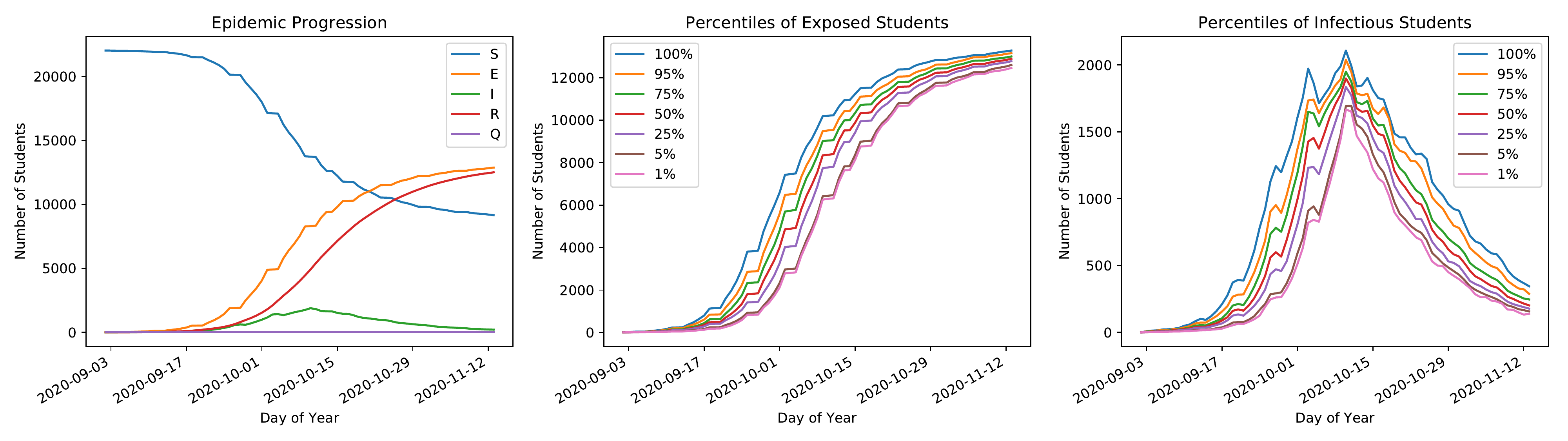}
    \caption{The three output charts from the basic simulation with the modification that social groups have been added via random social clusters.  These groups meet for 2 hours on weekdays and 3 hours on weekends.}
    \label{fig:randomsocial}
\end{figure}

Many of these function utilize randomization.  In order to ensure that a new randomized network is used with each simulation repetition, there is a special option in the \verb|Parameters| object to set a function \verb|preprocess|, which as argument takes a \verb|Semester| object and returns a new \verb|Semester| object.  This function is called to create a new \verb|Semester| object prior to each repetition.  Its usage is demonstrated in the following example, which uses the same definition of \verb|ClusterSettings| as above, but instead of creating a single \verb|Semester| object it creates a new version each time.  
\begin{lstlisting}
def groups_random(semester):
        clusters = make_randomized_clusters(semester.students, settings)
        return make_from_clusters(semester, clusters)

parameters = Parameters(reps = 100)
parameters.preprocess = groups_random
run_repetitions(semester, parameters)
\end{lstlisting}
Sample results are shown in Figure \ref{fig:randomsocial}.

\begin{figure}
    \centering
    \includegraphics[width=\textwidth]{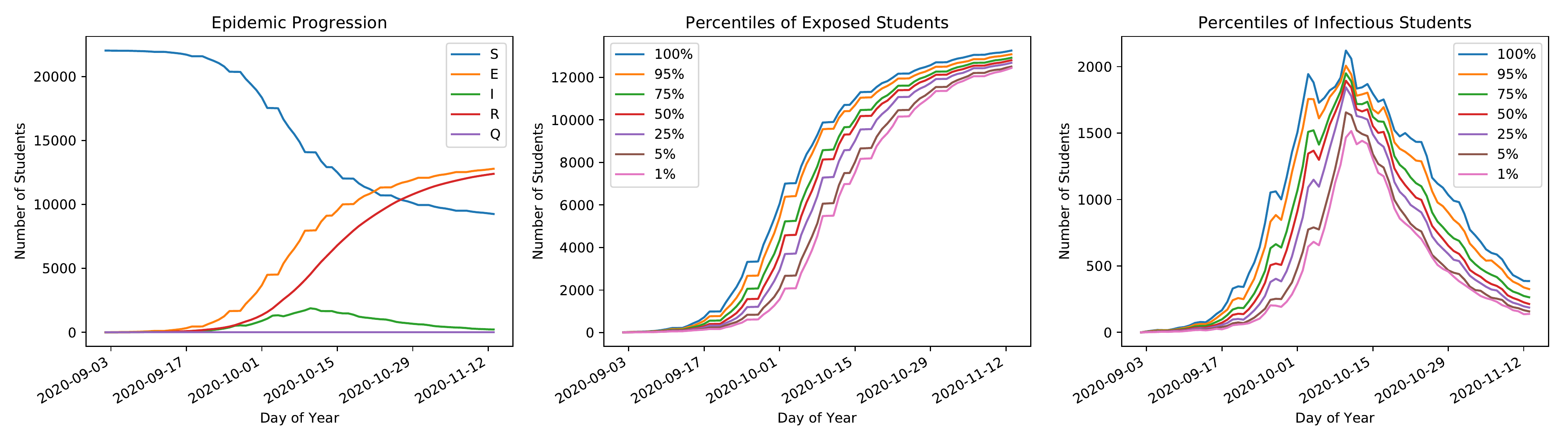}
    \caption{The three output charts for the basic simulation with the modification that 25\% of individuals have been randomly assigned to be pairs that meet for the equivalent of 20 hours each day, which accounts for hypothesized lack of social distance and not wearing face masks.}
    \label{fig:pairs}
\end{figure}

Similarly, we can use the \verb|preprocess| capability to create new sets of pairs of individuals in each repetition.
\begin{lstlisting}
def pairing(semester):
    clusters, _ = make_social_groups_pairs(semester, 0.25, interaction_time = 1200, weighted=False)
    return make_from_clusters(semester, clusters)

parameters = Parameters(reps = 10)
parameters.preprocess = pairing
run_repetitions(semester, parameters)
\end{lstlisting}
Sample results are shown in Figure \ref{fig:pairs}.

\subsection{Social meetings from groups.}

The final example we will give shows how to utilize group information loaded in the input file.  Recall that group information in the input file is optional but can be useful for creating custom groups.  Here we will make social groups corresponding to varsity athletic teams.  Within each of the varsity sports, each day we will split the group into clusters of six (to the extent possible) under the assumption that six is the maximum group size allowed to interact.  Individuals are assumed to interact with all other individuals within the cluster of six for 180 minutes per day.  Included in the data set that we loaded are entries like the following
\begin{lstlisting}
{'group': {'name': "Men's Varsity Squash", 'members': [170, 610, 630, 950, 970, 1150, 1350, 1530, 1590, 1630, 1770, 1850, 1870, 1890, 1910]}}
\end{lstlisting}
with we have created to mimic the composition of of each varsity sports team.

\begin{figure}
    \centering
    \includegraphics[width=\textwidth]{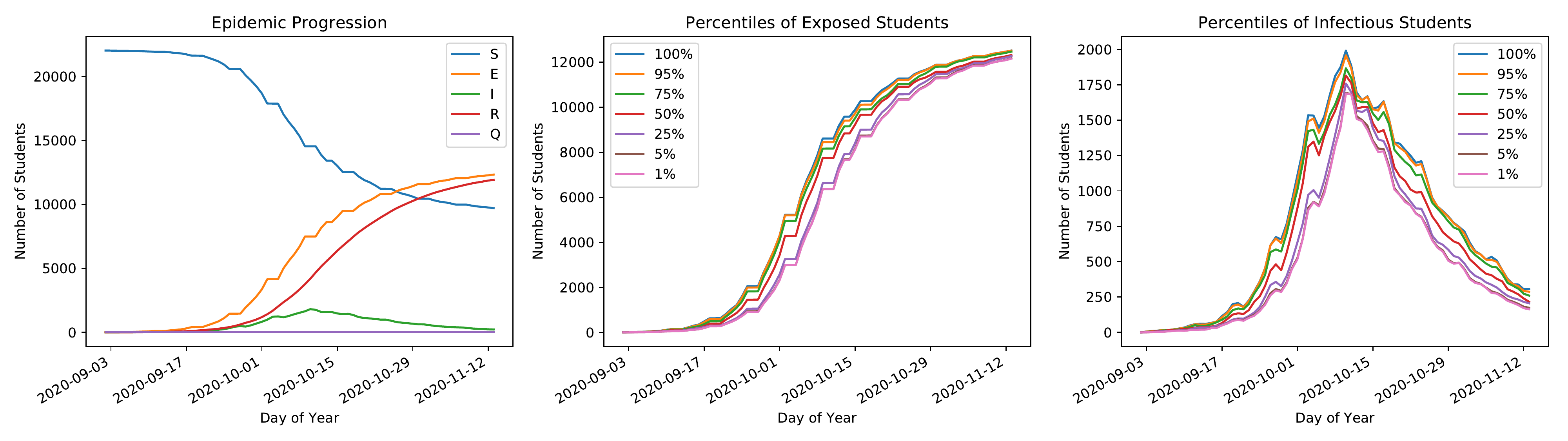}
    \caption{The three output charts for the basic simulation with the modification that members of varsity teams spend time with each other each day, with clusters changing within each team each day.}
    \label{fig:varsity}
\end{figure}

The built-in function we use to process these teams and create clusters is shown below.
\begin{lstlisting}
def make_social_groups_varsity(semester, 
                               weekday_size, weekday_time, 
                               weekend_size, weekend_time):
    varsity = {name : members for name, members in semester.groups.items()
               if 'Varsity' in name}
    max_size = max([len(x) for x in varsity.values()])
    cluster_settings = ClusterSettings(min(semester.meeting_dates), max(semester.meeting_dates),
                                       max_size, weekday_size, weekday_time,
                                       max_size, weekend_size, weekend_time)
    clusters = []
    processed_students = set()
    for sport, roster in varsity.items():
        clusters.extend(make_randomized_clusters(roster, cluster_settings))
        processed_students.update(roster)
    
    return clusters, processed_students
\end{lstlisting}
First the function reads through the list of groups in the \verb|Semester| object and saves those that have ``Varsity" in the name.  Since the behavior of the \verb|ClusterSettings| object is to make only a limited number of groups from the total population, we set its parameters \verb|weekday_group_count| and \verb|weekend_group_count| to a sufficiently large number (\verb|max_size|) so that the group count constraint never causes anyone on the roster to be omitted from placement in a cluster.  Then, for each sports team we use the \verb|ClusterSettings| object to make the clusters based on the team's roster.  We return the set of clusters as well as a set called \verb|processed_students| which let's the user know which individuals are present in clusters.  This may be useful if you don't want individuals assigned a group by one function to be assigned another group in a another function.  The following code runs the scenario with varsity student social groups.
\begin{lstlisting}
def groups_varsity(semester):
        clusters, processed = make_social_groups_varsity(semester, 6, 240, 6, 240)
        return make_from_clusters(semester, clusters)

parameters = Parameters(reps = 10)
parameters.preprocess = groups_varsity
run_repetitions(semester, parameters)
\end{lstlisting}
Sample results are shown in Figure \ref{fig:varsity}.

\begin{figure}
    \centering
    \includegraphics[width=\textwidth]{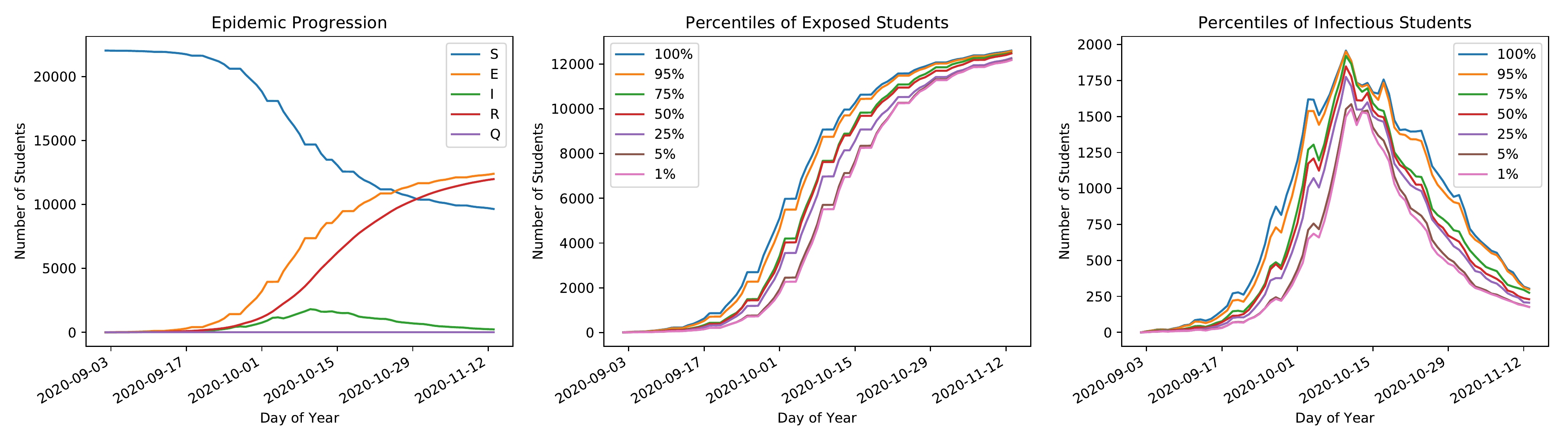}
    \caption{The three output charts for the basic simulation with the modification that members of varsity teams spend time with each other each day.  In this version, clusters do not change over the course of the semester.}
    \label{fig:varsity_static}
\end{figure}

Under some social models, the remixing of groups into new clusters each day may seem excessive.  In this case, we can substitute the function \verb|make_randomized_cluster| with \verb|make_randomized_static_clusters|, which takes the same arguments but uses the same clusters each day, defaulting to weekday settings for each day.  Demonstrating this alternative of the previous scenario, the following code produces the result shown in Figure \ref{fig:varsity_static}.
\begin{lstlisting}
def groups_static_varsity(semester):
        clusters, processed = make_social_groups_varsity_static(semester, 6, 240)
        return make_from_clusters(semester, clusters)

parameters = Parameters(reps = 10)
parameters.preprocess = groups_static_varsity
run_repetitions(semester, parameters)
\end{lstlisting}

\subsection {Overview of other \texttt{Parameters} settings}

Not all of the options available in the \verb|Parameters| object were described in the examples above.  This section lists and summarizes the members and how to use them.

\begin{itemize}
    \item  \verb|verbose| - Output day-by-day statistics of infection states, as well as computation time.
    \item  \verb|rate| -  Probability that an infectious individual spreads their infection to a susceptible individual per effective minute of contact time.
    \item  \verb|daily_spontaneous_prob| - Probability that a susceptible individual becomes infected each day from an external source that is not part of the simulation.  This models infections coming from outside the community.
    \item  \verb|contact_tracing| - Python object that has member function \verb|trace| that takes as arguments \verb|Simulator| and a date with  no return value.  It performs contact tracing as described in previous sections.
    \item  \verb|intervention_policy| - Python object that has member function \verb|testing| that takes as arguments \verb|Simulator| and a date with no return value.  It performs interventions, such as testing, as described in earlier sections.
    \item  \verb|infection_duration| - Python object that has member function \verb|duration| with arguments \verb|Simulator|, student id, and date.  It then returns the date the individual will become infectious, the day they will stop being infectious, and a Boolean indicating if this is because the individual will become symptomatic.
    \item  \verb|quarantine_length| - Default number of days individuals sent to quarantine spend before being returned to the general population, if they are in fact not infected.
    \item  \verb|preclass_interaction_time| - Extra time individuals spend near each other at the beginning and end of class that adds extra interaction time due to entering and leaving the classroom.  Only applies to meetings of type \verb|MeetType.COURSE|.
    \item  \verb|initial_exposure| - Either a list of individuals who are initially exposed or an integer for a number of individuals to draw at random at the beginning of the simulation that are exposed.
    \item  \verb|preprocess| - Function that takes \verb|Semester| object as input and output that performs any preparation, including randomization, prior to each repetition.
    \item  \verb|repetitions| - Number of simulation repetitions to perform.
    \item  \verb|start_date| - Python \verb|Datetime| object representing the first day to simulate.  The default value is September 2, 2020.
    \item  \verb|end_date| - Python \verb|Datetime| object representing the last day to simulate.  The default value is November 13, 2020, the day before Thanks Giving break.
\end{itemize}

\section{Conclusions}

We encourage everyone to to visit our Git repository at \url{https://github.com/MAS-Research/SEIR-Campus}.  It includes a copy of the SEIR-Campus package as well as a Jupyter notebook containing all of the examples included in this paper.  We hope that the package will be useful and remind users that the intention of the package is to be customized to fit your needs.  We encourage user comments and suggestions for additional features that may be useful for us to include in the future.

\bibliographystyle{acm}
\bibliography{refs.bib}

\end{document}